\documentclass[aps,prl,twocolumn, reprint,superscriptaddress]{revtex4-1}

\usepackage{graphicx}
\usepackage{amsmath}
\begin{document}

\title{Towards loss compensated and lasing terahertz metamaterials based on optically pumped graphene}

\author{P. Weis}

\affiliation{Department of Electrical and Computer Engineering and Research Center OPTIMAS, University of Kaiserslautern, 67663 Kaiserslautern, Germany}

\author{J. L. Garcia-Pomar}

\affiliation{Instituto de \'Optica, C.S.I.C., Serrano 121, 28006 Madrid, Spain}

\author{M. Rahm}
\email[]{marco.rahm@eit.uni-kl.de}

\affiliation{Department of Electrical and Computer Engineering and Research Center OPTIMAS, University of Kaiserslautern, 67663 Kaiserslautern, Germany}


\begin{abstract}
It is evidenced by numerical calculations that optically pumped graphene is suitable for compensating inherent loss in terahertz (THz) metamaterials. In a first step, the complex conductivity of graphene under optical pumping is calculated and the proper conditions for terahertz amplification in single layer graphene are determined. It is shown that amplification in graphene occurs for temperatures up to room temperature and for moderate pump intensities when pumped at a telecommunication wavelength $\lambda=1.5~\mathrm{\mu m}$. Furthermore, the amplification properties of graphene are evaluated and discussed at a temperature as low as $T=77~\mathrm{K}$ and a pump intensity $I=300~\mathrm{mW/mm^2}$ to investigate the coupling between graphene and a plasmonic split ring resonator (SRR) metamaterial. The contributions of ohmic and dielectric loss mechanisms are studied by full wave simulations. As a result, it is found that the loss of a split-ring resonator metamaterial can be compensated by optically stimulated amplification in graphene. Moreover, it is shown that a hybrid material consisting of asymmetric split-ring resonators and optically pumped graphene can exceed the laser threshold condition and can emit coherent THz radiation at minimum output power levels of $6 0~\mathrm{nW/mm^2}$. The use of optically pumped graphene is well suited for loss compensation in THz metamaterials and paves the way to new kinds of coherent THz sources.  
\end{abstract}

\maketitle

\section{Introduction}

In the last decade, metamaterials have intensively been subject to fundamental physical studies and now have been developed to an advanced level of maturity which opens up new exciting routes to industrial applications. In recent years, particular attention has been given to the interaction between metamaterials and other natural and artificial media following a three-fold aim: first to adaptively change the metamaterial properties by dynamic tuning of the underlying substrate properties \cite{Chen2008, Paul2009}, second to develop sensitive sensors by exploiting strong local field enhancement \cite{Liu2010, Weis2011, Reinhard2012, Dietze2013} and third to use active materials to compensate and overcompensate loss in the metamaterial structure \cite{Zheludev2008, Walther2010, Stockman2010, Xiao2010, Fang2011, Hess2012}, only to name few examples. Especially loss compensation and active metamaterials have moved into the focus of scientists since damping in the metal structures inherently limits their applicability. Yet, active metamaterials are not only discussed in the context of loss compensation but also was considered as a novel opportunity for the development of innovative light sources based on the concept of spasing lasers \cite{Noginov2009, Stockman2010, Fang2010} or lasing spasers \cite{Zheludev2008, Fietz2012, Tanaka2010}.

However, the quest for transferring these concepts to the terahertz (THz) frequency regime is hampered by the well-known lack of suitable gain media in this specific part of the electromagnetic spectrum. Some of available gain media exist in the gaseous phase but their small density conflicts with the strongly localized fields observed in metamaterials. THz gain in solids can only be achieved with additional effort as e.g.\ can be seen in the case of germanium where a strong magnetic field must be applied in order to observe amplification \cite{Chamberlin2003}. Artificial gain media as quantum dots or quantum cascade materials appear to be more promising in this respect and have been demonstrated to strongly interact with plasmonic structures, however the fabrication and handling of quantum cascade lasers are not trivial \cite{Belkin2009, Walther2010, Adams2010, Hinkov2013, Diao2013}.

Just in the last few years graphene appeared to be another promising medium for providing gain in a THz metamaterial. The benefit of graphene for the THz-technology has been shown in various ways. For example, several scientists used graphene to voltage tune or to optically tune metamaterials and thereby use them as broadband THz wave modulators \cite{Rodriguez2011, Tassin2012, Lee2012a, Rodriguez2012, Weis2012, Lee2013}. Moreover, the unique electromagnetic properties of graphene allow one to use it for the design and implementation of transformation optical devices \cite{Chen2011, Vakil2011, Sun2012}. A further aspect which is currently widely discussed is THz amplification in graphene. Rhyzii and Otsuji et al. \cite{Ryzhii2007, Rana2008, Dubinov2009, Ryzhii2009, Ryzhii2010, Dubinov2011, Takatsuka2012} suggested and demonstrated the phenomenon of negative conductivity in optically pumped graphene, which leads to an amplification of radiation in the THz-regime.
This is an interesting finding with the potential to dramatically change the state-of-the-art of THz-technology, which until the present day has significantly suffered from the deficit in high power radiation sources which presume the existence of suitable gain media. In this context, graphene offers the possibility to amplify THz-radiation at room temperature and can be either pumped optically or by current injection \cite{Ryzhii2011}.
Furthermore graphene is perfectly suited for being embedded in metamaterial structures due to its microscopic thickness. This is of particular importance since the electromagnetic fields in metamaterials are both strongly enhanced and strongly confined near the metal structure and thus only interact with the gain medium in a confined volume.
	
\begin{figure}[]
	\centering
	\includegraphics[width=0.7\columnwidth]{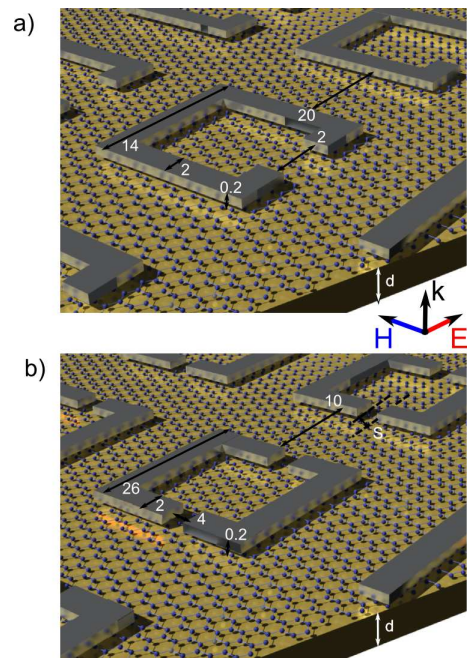}%
	\caption{Illustration of the (a) symmetric splitring resonator (SRR) and (b) the asymmetric splitring resonator (a-SRR) with dimensions indicated in $\mathrm{\mu m}$.}
	\label{fig:illustration}
\end{figure}

	Here, we numerically investigate amplification in a metamaterial with embedded graphene as a gain medium by means of 3-D full wave simulations. In a first step, we calculate gain in graphene based on a conductivity model that we derived from earlier work of Ryzhii et al. \cite{Ryzhii2007, Ryzhii2010, Karasawa2010} and determine the minimum optical power level that is required for obtaining amplification of THz waves in graphene. Second, we examine the interaction between metamaterials and optically pumped graphene based on two different types of plasmonic structures: a splitring resonator (SRR) and an asymmetric splitring resonator (a-SRR) metamaterial as illustrated in Fig.\,\ref{fig:illustration}. Equivalent for both structures we start with the strongly approximated case where we assume the absence of ohmic and dielectric loss. In a second approximation we take ohmic loss into account and finally also consider dielectric loss in a third approximation. That way we work our way along to realistic metamaterial/graphene composite structures. While the SRR/graphene composite material serves as an example for loss compensation in metamaterials whose resonators oscillate incoherently, the a-SRR/graphene metamaterial can be considered as a potential source for coherent, narrow-linewidth THz radiation complying with the concept of a lasing spaser \cite{Zheludev2008, Stockman2010}. Based on our numerical calculations we estimate that an optically pumped a-SRR/graphene metamaterial can provide coherent THz radiation of at least $60~\mathrm{nW}$ and thus can prospectively serve as an ultrathin, robust alternative to existing THz sources.

\section{Optically pumped graphene}
\begin{figure*}[]
	\centering
	\includegraphics[width=2\columnwidth]{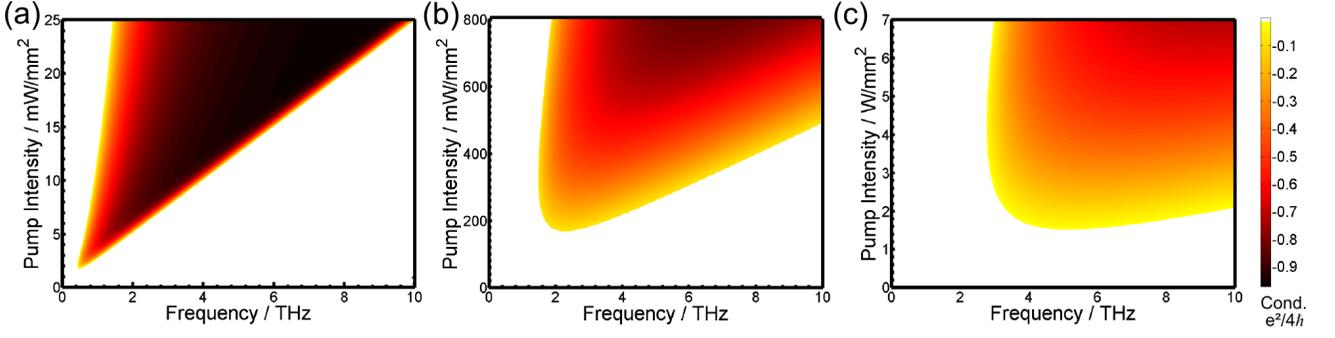}%
	\caption{Real part of the conductivity of graphene at a temperature of (a) T=4\,K,(b) T=77\,K and (c) T=300\,K. The conductivity is negative in the colored and positive in the white region.}
	\label{fig:Conductivity}
\end{figure*}

\begin{figure}[]
	\centering
	\includegraphics[width=0.7\columnwidth]{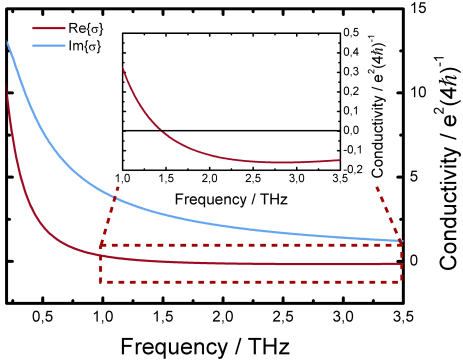}%
	\caption{\label{fig:Conductivity77K300mW} Real and imaginary part of conductivity for single layer graphene at T=77K 	$\tau_\mathrm{inter}=1~\mathrm{ns}$, $\tau_\mathrm{intra}=1~\mathrm{ps}$ and pumping intensity 	$I_\mathrm{pump}=300~\mathrm{mW/mm^2}$.}
\end{figure}

It was predicted by Rhyzii and Otsuji et al. in 2007 \cite{Ryzhii2007} that the conductivity of graphene can be driven to negative values by excitation of electron-hole pairs using an external light source in the visible or infrared spectral range. The efficiency for this excitation scheme is known to be $2.7~\%$. Immediately after pumping the electron-hole pairs relax via a fast channel into lower energy states. The relaxation process is mainly accompanied by the excitation of phonons of energy $\Delta E =0.2~\mathrm{eV}$ until the electron-hole pairs occupy an excited state with an energy below $\Delta E$. Note that although the energy of this populated state is $\leq 0.2$\,eV the energy level is above ground state. Since the relaxation time for these intraband transitions is of the order of $\tau\approx 1~\mathrm{ps}$ and therefore faster than the recombination time $\tau_\mathrm{r}\gtrsim 1~\mathrm{ns}$ for the electron-hole pairs, a population inversion can be obtained by this optical pumping scheme.

The optical generation of electron-hole pairs in graphene can be described by quasi-Fermi-levels for electrons $\mu_{\mathrm{Fe}}$ and holes $\mu_{\mathrm{Fh}}$ of the same absolute value $\mu_{\mathrm{Fe}}=-\mu_{\mathrm{Fh}}=\mu_{\mathrm{F}}$, so that the energy distribution of both, electrons and holes at temperature $T$ can be described by the Fermi distribution
\begin{equation}
	f(\epsilon)= \frac{1}{1+exp{\frac{\epsilon-\mu_\mathrm{F}}{k_B T}}}
\end{equation}
Hereby, $\epsilon$ denotes the energy of the considered electrons or holes and $k_B$ is the Boltzmann constant.

The complex intraband and interband conductivities $\sigma_{intra}$ and $\sigma_{inter}$ of single layer graphene experienced by an electromagnetic wave of frequency $\nu=\omega/2\pi$ can be described as \cite{Falkovsky2007a, Ryzhii2011}
\begin{eqnarray}
	\sigma_\mathrm{intra}& = &-\frac{e^2}{ \pi \hbar} \int_0^{\infty} \mathrm{d\epsilon} \frac{i \left|\epsilon\right|}{\hbar(\omega+i/\tau)} \cdot  \frac{\mathrm{d}f(\epsilon)}{\mathrm{d\epsilon}}	 \label{eqn:sigma_intra_1} \\
	\sigma_\mathrm{inter}& = &\frac{e^2}{4 \hbar} \left(1-2f(\frac{\hbar\omega}{2})-\frac{8\hbar\omega}{i\pi} \int\limits_{0}^\infty \mathrm{d\epsilon} \frac{f(\epsilon)-f(\frac{\hbar\omega}{2})}{(\hbar\omega)^2-4\epsilon^2}    \right) \label{eqn:sigma_inter_1}
\end{eqnarray}
where $\tau$ is the intra-band transition time and $e$ is the elementary charge.

By solving the integrals in (\ref{eqn:sigma_intra_1}) and (\ref{eqn:sigma_inter_1}) we can approximate the complex quantities of $\sigma_{intra}$ and $\sigma_{inter}$ and separate them into their real and imaginary parts:
\begin{align} 	
\sigma_\mathrm{intra} =& \frac{2e^2k_BT\tau}{\pi \hbar^2 (1+\omega^2\tau^2)} \cdot \log \left( 1+ \exp(\frac{\mu_\mathrm{F}}{k_BT}) \right)\label{eqn:sigmaIntra}\\
	&+i \frac{2e^2k_BT\omega}{\pi \hbar^2 (\omega^2+1/\tau^2)} \cdot \log \left( 1+ \exp(\frac{\mu_\mathrm{F}}{k_BT}) \right) \nonumber\\
	\sigma_\mathrm{inter} =& \frac{e^2}{4\hbar} \cdot \tanh \left( \frac{\hbar\omega-2\mu_\mathrm{F}}{4 k_B T} \right)\label{eqn:sigmaInter}\\
		&+i \frac{e^2}{8\hbar\pi} \cdot \log \left( \frac{(\hbar\omega+2\mu_\mathrm{F})^2}{(\hbar\omega)^2+(2k_B T)^2} \right) \nonumber
\end{align}
Following \cite{Ryzhii2007} we can describe the energy splitting of the  quasi-fermi levels $\mu_{\mathrm{F}}$ by
\begin{equation}
	\mu_{\mathrm{F}}=6\alpha\left(\frac{v_F}{k_BT}\right)^2 \frac{\hbar\tau_r}{\pi\nu} \cdot I
	\label{eqn:intensity}
\end{equation}
where $\alpha$ denotes the fine-structure constant, $v_F$ is the Fermi-velocity of charge carriers in graphene, $\tau_r$ is the recombination time for electron-hole pairs and $I$ describes the intensity of the photo-doping pump source.

As one can see from Eq.\ \ref{eqn:sigmaInter} the real part of the conductivity $\sigma_\mathrm{inter}$ can assume negative values when the quasi-Fermi level splitting becomes larger than the photon energy of the traveling wave through the material. The material enters this regime when it is pumped at high field intensities $I$ (see Eq. \ref{eqn:intensity}). Since the real part of $\sigma_\mathrm{inter}$ describes absorption of an electromagnetic wave in the material, a negative value of $\sigma_\mathrm{inter}$ corresponds to amplification of the wave in the medium. The imaginary part of the conductivity on the other hand allows one to calculate the phase difference between the electric field in the material and the exciting field and thus the retardation of the propagating electromagnetic waves. This phase information is of essential relevance for the design of a hybrid system consisting of a metamaterial and graphene since the interband conductivity of graphene remarkably offsets the resonance frequency of sub-wavelength resonators that interact with the amplifying medium.


Figures \ref{fig:Conductivity}(a)-(c) display the real part of the conductivity $\sigma$ for $\tau=1~\mathrm{ps}$, $\tau_r=1~\mathrm{ns}$, $\lambda=1.5~\mathrm{\mu m}$ at liquid helium temperature $T=4~\mathrm{K}$, liquid nitrogen temperature $T=77~\mathrm{K}$ and room temperature $T=300~\mathrm{K}$ as derived from Eqs. \ref{eqn:sigmaIntra} -- \ref{eqn:intensity}. The colored regions indicate regimes where the frequency and the pump intensities are chosen such that graphene provides gain and amplifies a wave traveling through the material. As can be seen, we obtained amplifying regimes at all investigated temperatures. For increasing temperature the frequencies where amplification can be observed shift to higher values while at the same time the required pump threshold intensity for amplification increases.
Although graphene at temperatures near $T=0$\,K seems to be most beneficial for amplification of THz waves, we restrict ourselves to the investigation of hybrid metamaterial/graphene composites at liquid nitrogen temperatures of $T=77~\mathrm{K}$ to ensure practical usefulness. The minimum pump intensity at $77~\mathrm{K}$ is $I=162~\mathrm{mW}$. In this temperature range we can exploit the benefits of moderate pump intensities and readily available, straightforward cooling procedures. At this point we should note that operation of metamaterial/graphene composites with amplification at room temperature also appears to be feasible despite the demand for higher pump intensities which, however, can be obtained by use of commercial laser systems.

  \section{Loss compensated and lasing metamaterials}

In this section we investigate hybrid systems of optically pumped graphene and metamaterials in the THz-regime. In the following considerations we targeted a high amplification in a frequency range between 2 and 3\,THz. At the same time we strived for a pump intensity regime below $I=300$\,mW$/$mm$^2$.
The electromagnetic interaction between a 2-dimensional array of micro-resonators and graphene is a sophisticated process that strongly depends on the electric field distribution around the micro-resonators and the penetration into graphene. In this context, analytical methods cannot describe all phenomena contributing to this coupling process and a numerical approach becomes inevitable. For this purpose, we numerically calculated the electromagnetic fields in the metamaterial/graphene composite by use of the frequency solver of CST Microwave Studios.

In a first step, we implemented the graphene layer as an infinitesimally thin sheet with a surface impedance $Z=1/\sigma$ and a conductivity $\sigma$ that we calculated based on Eqs.\ \ref{eqn:sigmaIntra} and \ref{eqn:sigmaInter}. For all of the following numerical calculations we assumed a pump intensity $I=300~\mathrm{mW/mm^2}$ at a temperature of $T=77~\mathrm{K}$. The resulting real and imaginary parts of the graphene conductivity are depicted in Fig.\ \ref{fig:Conductivity77K300mW}. The real part of the conductivity is negative at frequencies above $1.5~\mathrm{THz}$ and reaches a minimum of $\mathrm{Re}(\sigma) = -0.16\cdot e^2/4\hbar$ at $\nu_0\approx 2.8~\mathrm{THz}$, which corresponds to a maximum of amplification. For this reason, we designed the metamaterials to operate close to frequency $\mu_0$. Moreover, we presumed that the amplification in graphene was constant and independent from the intensity of the transmitted THz wave. Strictly speaking this assumption only holds for low input intensities of the incident THz wave which is discussed in more detail in section 4. To test the reliability of the numerically calculated THz wave transmission through and reflection from a single-layer graphene sheet we compared the numerical data to analytically obtained values derived from the Fresnel equations and found good agreement \cite{Stauber2008}. In a further step we implemented the metamaterial structure on top of the single-layer graphene. In a first approach we assumed that the metamaterial structure was composed a $200~\mathrm{nm}$ thick perfect electrical conductor (PEC). In a second step we investigated how the use of a lossy metal as metamaterial structure influences the amplification process in a metamaterial/graphene hybrid. Finally, we examined gain in such a composite when a lossy substrate is added and thus iteratively approached real-world conditions. It should be noted that no graphene was present directly under the metal structure of the metamaterial for two reasons:  First, in the most common configuration the pump beam is incident from the metamaterial side, which inherently shadows the region under the metal structure such that this shadowed region should not play an immediate role in the amplification process in graphene and second an electric contact between the metal structure and graphene strongly influences the band potentials in graphene and most likely causes loss of gain in the graphene under the metal. Keeping these presumptions in mind, we first numerically investigate amplification in a hybrid metamaterial/graphene composite in the following section.

\subsection{Split-ring resonator}
Split-ring resonator (SRR) based metamaterials are one of the first suggested artificial media with magnetic response at high frequencies and ever since have been used in the design of a great variety of artificial optical components. It is well known that the magnetic resonance of SRR-based metamaterials can be excited either by the magnetic component of the electromagnetic field or by capacitive coupling of the electric component of the electromagnetic field due to the bi-anisotropy of the SRR. Here, we restrict our investigation to the case of an electromagnetic wave that is incident normal to the SRR plane and whose electric field is parallel to the split. That way the excitation of the resonance as well as the interaction with optically pumped graphene is mediated solely by the electric field.
\begin{figure*}[]
	\centering
	\includegraphics[width=1.6\columnwidth]{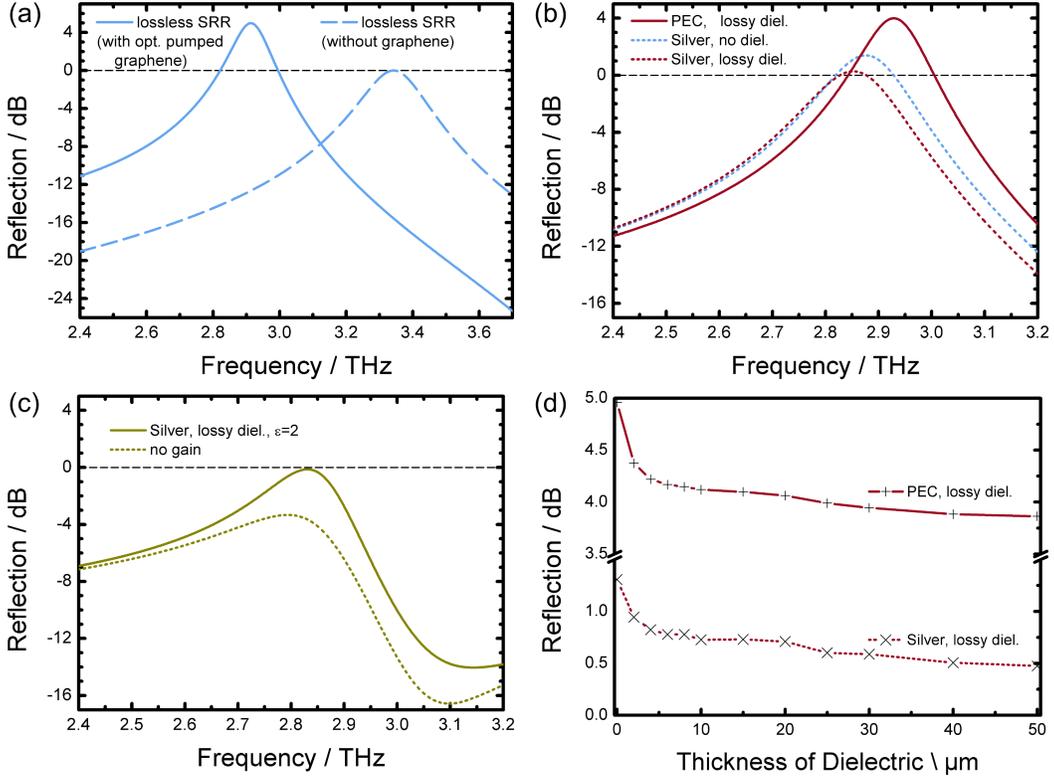}
	\caption{Reflectivity of optically pumped SRR/graphene metamaterial in the case of (a) no loss (b) ohmic and dielectric loss. The pump intensity was $I_{pump}=300$\,mW$/mm^2$ at $\lambda=1.5~\mathrm{\mu m}$. (c) Reflictivity of optically pumped SRR/graphene with a silver metamaterial structure on a substrate with a permittivity of $\epsilon=2$ and a loss tangent $\mathrm{tan}\delta=0.01$. (d) Maximal reflectivity of an optically pumped SRR/graphene metamaterial in dependence on the thickness of a lossy substrate with a permittivity of $\epsilon=1$. }
	\label{fig:SRRreflection}
\end{figure*}
	
The SRR functionality can be understood in analogy to the physical processes in a laser cavity. When an electromagnetic wave is incident on the SRR metamaterial it can excite oscillations in the SRRs that are resonant at the resonance frequency of the SRRs. This implies that the energy of the incident wave is transferred into oscillation energy of the SRRs. During oscillation the physical form in which the energy is stored in the SRRs alternates between magnetically stored energy which is related to the induced coil currents in the SRRs and electrically stored energy that is associated with the electric field in the capacitive gap of the SRRs. During each oscillation cycle the electric field in the capacitive gap is amplified by the underlying optically pumped graphene layer. A fraction of the stored energy in the SRRs is radiated by the structure in forward, but mainly in backward direction with respect to the incident wave. The radiative loss results in a damping of the SRR oscillation if the radiated energy exceeds the gain in graphene. Since a metamaterial consists of a whole array of oscillating SRRs that act as microresonators analogue to a laser cavity, it is important to note that these microresonators do not necessarily oscillate in phase. In fact, it is known that the oscillations in an array of SRRs do not couple coherently which means that the resonant oscillations of neighboring SRRs are not strictly in phase. In consequence, an SRR based metamaterial coupled to optically pumped graphene is only expected to offset damping loss in the metamaterial, yet cannot be used as a coherent radiation source.
\subsubsection{SRR-based metamaterial without ohmic or dielectric loss}
In a first approach, we consider a free-standing, substrate-free SRR-based metamaterial without ohmic loss coupled to an optically pumped single-layer graphene sheet. Although we are well aware that a free-standing, substrate-free SRR-based metamaterial on graphene cannot exist in reality due to the lack of mechanical stability, it is instructive to study such a hypothetical material in a first step due to a strongly reduced complexity of the investigated medium. In order to study amplification in an optically pumped SRR/graphene metamaterial it is instructive to investigate the reflectivity and transmissivity of such a medium. Since an SRR metamaterial acts as a bandstop filter around resonance frequency in transmission, it is more compelling to investigate only the reflectivity of the material. In this respect, we calculated the reflection from the SRR/graphene hybrid medium when it is pumped at an intensity of $I=300~\mathrm{mW/mm^2}$. The maximal reflectivity of optically pumped SRR/graphene reaches $4.92~\mathrm{dB}$ which evidences amplification in the hybrid SRR/graphene medium as indicated in Fig.\ \ref{fig:SRRreflection}(a) (solid curve). For reference, a reflectivity of $0~\mathrm{dB}$ corresponds to the case that all the energy of an incident wave is reflected back. In this respect, a reflectivity of $4.92~\mathrm{dB}$ indicates amplification in the hybrid SRR/graphene medium. To allow a better classification of these results we additionally plotted the reflectivity of the same metamaterial in absence of the graphene layer. Comparing both curves implicates two conclusions: 1. Resonant gain in a SRR/graphene metamaterial significantly exceeds the threshold for loss compensation. 2. The imaginary part of the graphene conductivity strongly affects the resonance frequency of the metamaterial and requires consideration in hybrid metamaterial/graphene components.
\subsubsection{SRR-based metamaterial with ohmic loss}
Evidently metamaterials are inherently lossy. At least ohmic loss in the conductor cannot be avoided and demands consideration. In the THz regime, ohmic loss is significantly lower than in the optical frequency spectrum and can be adequately expressed in terms of the conductivity of the metal. We calculated the conductivity of different metals (gold, copper and silver) and summarized the results in Tab.\ \ref{tab:SRRConductivity}. Furthermore we calculated the maximum reflectivity that we expect from an SRR/graphene hybrid material when the different metals are used. For example, we depicted the reflectivity spectrum of silver, shown as curve with the label ''silver, no diel.'', in Fig.\,\ref{fig:SRRreflection}(b). As can be seen by comparison of the reflection spectrum of an SRR/graphene medium with PEC resonators and silver resonators, the resonance frequency of the SRR/graphene medium decreases with the introduction of loss. At the same time the reflectivity sharply drops with the introduction of metal loss and thus behaves inverse to the conductivity. In the THz regime a great variety of metals are well suited for being used for the implementation of hybrid graphene/SRR metamaterials, since the conductivities of most metals are high and of comparable magnitude in this frequency range of interest. Although loss is crucial for the functionality of the metamaterial, our numerical calculations evidenced that the reflectivity of an optically pumped SRR/graphene hybrid only differs slightly dependent on the choice of silver or copper as a plasmonic component (see Tab.\ \ref{tab:SRRConductivity}). Yet, we also observed that gold is not the metal of choice in order to obtain strongly amplified reflectivity. The reason for this lies in the reduced conductivity of gold in comparison with silver and copper.
 \begin{table}[]
 \caption{Reflection at resonance frequency in dependence of the SRR's material.}
 \centering
	 \begin{tabular}{@{}lcc@{}}
	 	\hline
  		Metal  & Conductivity S/m~ 	 & ~Max. refl. /dB \\ 
  		\hline
  		PEC    & $\infty$ 	 		 & 4.98 \\				   	
  		Silver & $6.30\cdot10^7$ 	 & 1.40 \\					
  		Copper & $5.96\cdot10^7$ 	 & 1.32 \\					
  		Gold   & $4.56\cdot10^7$ 	 & 0.94 \\					
  		\hline
 	\end{tabular}
 	
 	\label{tab:SRRConductivity}
 \end{table}
\subsubsection{SRR-based metamaterial with dielectric loss}
	In the following we discuss the impact of dielectric loss on the amplification in an optically pumped SRR/graphene system. Since substrate-less metamaterials do not provide the necessary mechanical stability, the use of a substrate seems to be compelling. However, any (dielectric) substrate inherently introduces additional loss into the hybrid system. The requirements on such a substrate material include mechanical stability, a high optical damage threshold, good metal adhesion properties and low electromagnetic loss in the THz region. Usually THz metamaterials are fabricated e.g. on polimide \cite{Lee2012a}, PDMS \cite{Khodasevych2012} or benzocyclotene \cite{Weis2011}, which are materials that meet the foregoing requirements. To model the impact of a dielectric substrate on the electromagnetic properties of an SRR/graphene composite we assumed a material of thickness $d$ with a constant loss tangent $\tan\delta=0.01$, For the sake of comparability we chose the permittivity $\epsilon=1$ in our preliminary studies. We are well aware that dielectric media with such a low permittivity do not naturally exist, yet it is instructive to first investigate such a simplified material system for understanding amplification in composite SRR/graphene media. We will release this constraint in the course of the discussion to account for real-world dielectric substrates. In all calculations we assumed the dielectric substrate in contact with the graphene underneath the metal structure.

Figure \ref{fig:SRRreflection}(b) shows the reflectivity of an optically pumped SRR/graphene metamaterial with a dielectric substrate of thickness $d=10~\mathrm{\mu m}$. As a metal for the SRR structure we presumed a perfect electric conductor (PEC). The contribution of the dielectric loss to the total loss in this case is smaller than the fraction of typical ohmic loss in a real metal (see foregoing discussion), but still significant. Furthermore, we observed that the amplification of the reflected wave is stronger when substrates of small thickness are used and that amplification decreases for increasing substrate thickness as can be seen in Fig.\,\ref{fig:SRRreflection}(d). In this context, the amplification factor becomes small for thicknesses larger than $25~\mathrm{\mu m}$. The saturation behavior reflects the local electric field distribution of the SRR, which is strongly localized near the metamaterial plane. As a consequence, the SRRs interact strongly with the nearby graphene layer which is the source of amplification in the composite material. In contrast, they almost do not interact with the lossy substrate located further away from the SRRs. This leads to the exceptional situation, that one can obtain amplification in an active SRR/graphene metamaterial although the substrate loss surpasses the gain in graphene. This can be easily explained by taking the decisive role of the geometrical distance between the amplifying graphene and the SRRs and respectively between the lossy substrate and the SRRs into account.
\subsubsection{SRR-based metamaterial with ohmic and dielectric loss}	
In the foregoing discussion we learned that ohmic loss in the metallic metamaterial structure is significantly higher than dielectric loss in the used substrate as can be seen in Fig.\,\ref{fig:SRRreflection}(b). For this reason, the reflectivity of an SRR/graphene composite with pure ohmic loss is very similar to the reflectivity of an SRR/graphene composite with ohmic and dielectric loss.  As illustrated in Fig.\,\ref{fig:SRRreflection}(b), the maximum reflectivity of optically pumped SRR/graphene still exceeds $0.28~\mathrm{dB}$ when both dielectric and ohmic loss are taken into account. A reflectivity higher than $0~\mathrm{dB}$ is a clear indication for amplification in the active composite metamaterial.

As discussed before, we have to select the permittivity of the substrate to be $\epsilon\neq 1$ for being more realistic in our model. In the further discussion, we exemplarily chose $\epsilon= 2$ which corresponds to a typical value found in dielectrics at THz frequencies. At this point it should be noted that the change of substrate permittivity inherently results in a shift of the SRR resonance frequency. In order to maintain comparability with earlier results, we adapted the arm length of the SRRs to $12.8~\mathrm{\mu m}$ to restore the resonance frequency of the SRRs. That way we could rule out an impact of a resonance frequency shift on the amplification factor, however had to pay the cost that we could not keep the resonator quality of the SRRs unchanged which certainly also affects the amplification process. The reflectivity of this realistic metamaterial/graphene system is depicted in Figure \ref{fig:SRRreflection} (c). Under these conditions, we obtained a reflectivity of the optically pumped SRR/graphene metamaterial of  $-0.1~\mathrm{dB}$ at resonance. The reflectivity was very close to $0~\mathrm{dB}$ and thus almost equaled the case of exact loss compensation. To compare the reflectivity of optically pumped SRR/graphene with the corresponding unpumped SRR/graphene metamaterial we set the real part of the graphene conductivity to $\mathrm{Re}\left\lbrace \sigma \right\rbrace=0~\mathrm{S/m}$ while preserving the imaginary part. Strictly speaking, the imaginary part of the graphene conductivity also has to change slightly in the unpumped regime, however, this change is very small and can be neglected. In good approximation we can therefore assume that the resonance frequency remains unchanged for the unpumped SRR/graphene metamaterial which eases comparability. Figure \ref{fig:SRRreflection}(c) shows the reflectivity of the passive hybrid metamaterial. As can be seen, the reflectivity is $-3.3~\mathrm{dB}$ and, as expected, significantly smaller than in the loss-compensated case of corresponding optically pumped SRR/graphene. Hence, we obtained a gain of $3.2~\mathrm{dB}$ in optically pumped SRR/graphene.
\subsection{Asymmetric split-ring resonator}
	In the following we discuss amplification in an optically pumped hybrid metamaterial consisting of asymmetric splitring resonators (a-SRRs) and graphene as depicted in Fig.\,\ref{fig:illustration}(b). As before, the optical pump intensity was $I=300~\mathrm{mW/mm^2}$. The a-SRR differs fundamentally from the ordinary splitring resonator with a single split since it oscillates in a so-called dark mode. If the position of both splits of the SRR lies in the symmetry plane of the SRR, it is not possible to excite the SRR by external electromagnetic dipole radiation. Reversely, assuming oscillations in the symmetric double split ring resonator, no radiation is coupled out of the SRR in this symmetric configuration which is the reason for terming the mode of the oscillations as a dark mode. In order to excite oscillations the symmetric arrangement of the split gaps must be broken. In our case we achieved this by shifting both splits in the same direction away from the symmetry plane of the double split ring resonator, thus obtaining an a-SRR. In order to quantitatively describe the magnitude of the introduced asymmetry, we defined a symmetry factor $S$ which describes the part of split length which is shifted beyond the position of the symmetry plane. In this respect a value of $S=0$ describes a perfectly symmetric position of both splits, while $S=1$ means that the splits are shifted beyond the symmetry plane in their full length as depicted in the inset of Fig.\,\ref{fig:ASRRTransmission}(a). As the asymmetry is increased the resonator quality factor decreases and the resonator bandwidth gets broader. At the same time the radiation loss of the resonator increases for higher asymmetry of the a-SRRs. At this point it becomes clear that high transmission through the metamaterial at resonance frequency can only be achieved by a trade off between a sufficiently high quality factor of the a-SRRs and sufficiently high emission of radiation from the a-SRRs. We observed that the optimal value of asymmetry $S$ strongly correlates with the amount of loss that is inherent in the system. Especially ohmic as well as dielectric loss come along with a reduction of resonator quality and have to be taken into account for the optimization process. 		
\subsubsection{a-SRR-based metamaterial without ohmic or dielectric loss}
\begin{figure}
	\centering
	\includegraphics[width=0.8\columnwidth]{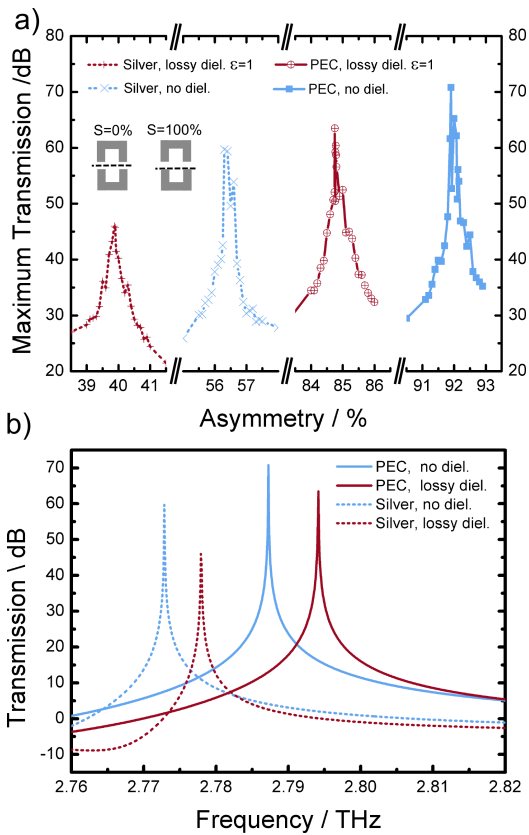}
	\caption{(a) Maximum of the amplitude transmission in dependence of the asymmetry factor, (b) Transmission spectrum of the a-SRRs for different cases: 1. lossless (blue solid curve), 2. ohmic loss in silver (blue dashed curve), 3. dielectric loss (red solid curve), 4. ohmic loss in silver and dielectric loss in the substrate (red dashed curve).}
	\label{fig:ASRRTransmission}
\end{figure}

	In a first approach we restrict ourselves to the consideration of a loss-less a-SRR/graphene hybrid metamaterial whose metallic structures are composed of PECs. As in the foregoing discussion of the SRR/graphene metamaterials, we first hypothetically assume the metamaterial to be substrate-less. Figure\,\ref{fig:ASRRTransmission}(a) illustrates the dependence of the maximal amplitude transmission through the optically pumped a-SRR/graphene metamaterial from the asymmetry factor $S$. The blue solid curve corresponds to the investigated case of an optically pumped loss-less a-SRR/graphene hybrid metamaterial. We found that the maximal transmission was obtained for $S\approx 91.9~\mathrm{\%}$. The resonance frequency was $\nu_0=2.78~\mathrm{THz}$. As expected, Fig.\,\ref{fig:ASRRTransmission}(a) furthermore evidences a strong dependence of the maximal transmission through the metamaterial on the asymmetry factor $S$. The full width at half maximum of the maximal transmission dependent on the asymmetry factor $S$ is as low as $ \Delta S\approx 1.6~\mathrm{\%}$, which corresponds to a shift of the split by only $15~ \mathrm{nm}$. This shift is by about 3 orders of magnitude smaller than the geometrical size of the a-SRR unit cell which is ($36~\mathrm{\mu m}$) and therefore imposes incredibly high accuracy demands on the fabrication process.

Figure \,\ref{fig:ASRRTransmission}(b) shows the amplitude transmission spectrum through an optically pumped loss-less a-SRR/graphene hybrid metamaterial (blue solid curve) for an optimized asymmetry factor $S$. We observe that the amplification of transmitted THz radiation through the a-SRR/graphene hybrid metamaterial at resonance frequency is remarkably higher than the amplification of reflected THz radiation from an optically pumped loss-less SRR/graphene hybrid metamaterial. This can be understood in terms of the higher resonator quality of the a-SRR metamaterial structure and consequently the narrower resonator bandwidth of the a-SRRs compared to the SRRs. Another important aspect of a-SRR structures, which is not further addressed in this work, is that neighboring a-SRRs coherently couple via the electric fields in the splits and thus oscillate in phase. For this reason a-SRR metamaterials cannot only be used to amplify radiation, but also may serve as a source of coherent radiation \cite{Zheludev2008, Fang2010}.
\subsubsection{a-SRR-based metamaterial with ohmic loss}
	Following along the lines of the previous discussion for SRR/graphene metamaterials we step-wise introduce different origins of loss in our investigation of an optically pumped a-SRR/graphene metamaterial. In a first step we consider ohmic loss by substituting the metallic PEC metamaterial structure by a silver structure. Due to a higher damping of the resonant oscillation ohmic loss results in a decrease of the resonator quality. As a result, the oscillation bandwidth of the a-SRRs gets broader and the energy from the a-SRRs is emitted into a wider spectral band such that the peak amplitude of the irradiated THz wave decreases. In order to optimize the peak electric field amplitude of the emitted THz radiation, we have to re-balance the trade off between resonator quality (and thus resonator bandwidth) and the out-coupling efficiency of the a-SRRs. As we learned in the earlier discussion, we can achieve this by re-optimizing the asymmetry factor $S$. As can be seen in Fig.\,\ref{fig:ASRRTransmission}(a) from the blue dashed curve, the optimal asymmetry factor $S$ shifts to $S=56.3~\mathrm{\%}$ compared to an asymmetry factor of $S= 91.9~\mathrm{\%}$ for the loss-less case. The necessity for changing the asymmetry factor $S$ in order to optimize the peak amplitude of the emitted radiation for the silver structure is a straightforward consequence of the fact that ohmic loss is a major loss channel in plasmonic metamaterials.

%
%
\subsubsection{a-SRR-based metamaterial with dielectric loss}
	As a second source for loss we consider a dielectric substrate of thickness $d=10~\mathrm{\mu m}$, $\tan\delta=0.01$, and $\epsilon=1$ underneath the graphene layer. The a-SRRs are composed of PEC. Again the asymmetry factor $S$ of the splits requires re-adjustment in order to optimize the peak amplitude amplification in the optically pumped a-SRR/graphene metamaterial. It should be noted that the adjustment of the asymmetry factor $S$ is not only dependent on the dielectric loss but also on the thickness of the dielectric. In this context, Fig.\,\ref{fig:ASRRTransmission}(a) displays the dependence of the maximal amplitude transmission through the metamaterial on the asymmetry factor $S$ for a $d=10~\mathrm{\mu m}$ thick substrate. We found the optimal asymmetry value to be $S=84.75~\mathrm{\%}$. Comparing the impact of pure dielectric loss on the amplification in an optically pumped a-SRR/graphene metamaterial with PEC resonators to the influence of pure ohmic loss in such an optically pumped a-SRR/graphene material without dielectric substrate (see foregoing section) we found that the dielectric loss has only a weaker degradation effect on the resonator quality (see Fig.\,\ref{fig:ASRRTransmission}(b)). In consequence, only small re-adjustments of the asymmetry factor $S$ were necessary in comparison with the loss-less case to re-optimize amplification.

	\subsubsection{a-SRR-based metamaterial with ohmic and dielectric loss}
	\begin{figure}
	\centering
	\includegraphics[width=0.8\columnwidth]{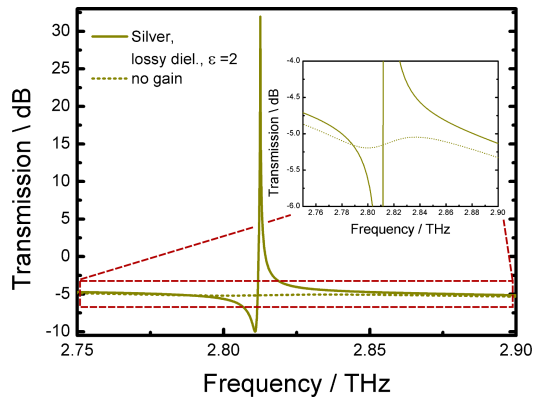}
	\caption{Transmission spectrum of the a-SRRs for ohmic loss in silver and dielectric loss in the substrate ($\epsilon=2$, $\tan\delta=0.01$).}
	\label{fig:Eps2}
\end{figure}
	The presence of both ohmic and dielectric loss in the material system the compared further reduces amplification in an optically pumped a-SRR/graphene composite metamaterial. As can be seen in Fig.\,\ref{fig:ASRRTransmission}(a) the asymmetry factor again requires re-adjustment to $S=39.875~\mathrm{\%}$ in order to balance the decrease of resonator quality caused by loss.

In a last step of our analysis we perform numerical full wave simulations on a real-world metamaterial composed of silver and a dielectric subtrate with a permittivity of $\epsilon=2$, as depicted in Fig.\,\ref{fig:Eps2}. Again we optimize the asymmetry factor of the a-SRR structure to maximize the transmission through the optically pumped a-SRR/graphene metamaterial. In this respect we determined a maximal transmission of $32~\mathrm{dB}$ for an optimized asymmetry factor of $S=13.125~\mathrm{\%}$ and a resonator arm length of $l=23~\mathrm{\mu m}$. In comparison with a transmission of $-5~\mathrm{dB}$ for the unpumped composite with $\mathrm{Re}\left\lbrace \sigma \right\rbrace=0~\mathrm{S/m}$ and preserved imaginary part this corresponds to a gain of $37~\mathrm{dB}$ at resonance. It is obvious that the amplification in such a real-world metamaterial is strongly reduced in comparison with the more idealized material systems that we discussed above. Yet, even when ohmic and dielectric loss of a substrate with permittivity unequal to one is considered we still observe that such composite structures can exceed loss compensation and thus can act as an active metamaterial. Thus, media composed of metamaterial structures that are embedded in a graphene environment potentially may serve as optically pumped lasing spasers that emit coherent radiation. Especially for the THz frequency range, where we still lack of powerful radiation sources, a-SRR/graphene composites might perform as active THz antennas and thus provide new ways for the generation of coherent THz waves.
\section{Constant gain restriction}
In this work we restricted ourselves to the case of constant gain. This simplification reduces the numerical complexity and increases calculation speed, thereby allowing a detailed examination of the structural properties. As a drawback the obtained results are only valid in a regime for which the excitation rate of electron-hole pairs is significantly higher than the rate of stimulated emission for the transition from the excited level to the ground state. In other words we assumed in our numerical calculations the population inversion in graphene to be constant. To determine the confidence interval for this assumption it is necessary to give an estimation for the relevant transition rates that we expect in a metamaterial/graphene medium.

For this purpose, we first consider the transition rate for electron-hole pairs in optically pumped graphene. The intensity of the incident pump light can be expressed in terms of photon energy by
\begin{equation}
I_{pump}=[n_{pump}\cdot c] \cdot h\nu_{pump}
\end{equation}
where $h$ is the Planck constant, $c$ is the light velocity, $\nu$ is the frequency of the pump wave and $n_{pump}$ is the volume density of incident pump photons. In this expression, the term $\tilde{n}_{pump}=n_{pump} \cdot c$ describes the number of pump photons per second that impinge onto an area of 1\,$m^2$. Furthermore, the intensity of the incident pump wave can be calculated in the wave picture as
\begin{equation}
I=\epsilon_0 c |\vec{E}_{pump}|^2
\end{equation}
where $\epsilon_0$ is the dielectric constant. In consequence, we obtain for the number of incident pump photons per second and square meter
\begin{equation}
\tilde{n}_{pump}=\epsilon_0 c |\vec{E}_{pump}|^2/(h\nu)=I_{pump}/(h\nu_{pump})
\end{equation}
We know from experiments that only a fraction of $2.7 ~\%$ of the incident pump photons is absorbed by the graphene so that we obtain a number of
\begin{equation}
\tilde{n}^{abs}_{pump}=0.027 \cdot \epsilon_0 c |\vec{E}_{pump}|^2/(h\nu)=0.027 \cdot I_{pump}/(h\nu_{pump})
\end{equation}
absorbed photons per (second$\cdot$m$^2$). Only these absorbed photons contribute to the population inversion.

The second question relates to the rate of THz photons that take part in the stimulated emission process in the metamaterial/graphene composite. From numerical calculations we estimated that only 0.74\% of THz photons that pass the optically pumped graphene participate in the amplification process and are responsible for a possible pump depletion. Furthermore, the THz electric field can experience local plasmonic field enhancement due to the metamaterial structure for which we account by introducing a spatially dependent enhancement factor $g=\frac{|\vec{E}_{THz,loc}|}{|\vec{E}_{THz}|}$. By this means we can calculate the fraction of photons per (second$\cdot$m$^2$) that deplete the inversion as
\begin{equation}
\tilde{n}_{THz}=0.0074\cdot \epsilon_0 c g^2 \frac{|\vec{E}_{THz}|^2}{h\nu_{THz}}=0.0074 g^2 \frac{I_{THz}}{h\nu_{THz}})
\end{equation}
In order to ensure operation in the linear, constant gain regime we only allow a maximum of 50\% of the upper state population to be depleted by stimulated emission. Presuming this condition we can rule out gain saturation in the active medium. In quantitative terms this requirement can be expressed by the condition $\tilde{n}_{THz}\leq 0.5 \tilde{n}^{abs}_{pump}$ which implies that the maximum amplitude of the incident THz electric field that is amplified in the metamaterial/graphene composite must be
\begin{equation}
|\vec{E}_{THz}| \leq \sqrt{0.5\cdot 3.65}\sqrt{\frac{\nu_{THz}}{\nu_{pump}}}/g|\vec{E}_{pump}|
\end{equation}
which is equivalent to the condition that the intensity of the incident THz wave is
\begin{equation}
I_{THz} \leq 0.5 \cdot 3.65/(g^2) \frac{\nu_{THz}}{\nu_{pump}} I_{pump} \label{eqn:I_THz}
\end{equation}
At this point we note that we performed the foregoing numerical analysis of metamaterial/graphene composites for an optical pump intensity of $I=300~\mathrm{mW/mm^2}$. Calculating the maximally allowed intensity of the THz wave for which operation in the constant gain regime can be expected we obtain $I_{THz} \approx 200~\mathrm{nW }$ by use of Eq.\ \ref{eqn:I_THz}.



\begin{figure}
	\centering
	\includegraphics[width=0.8\columnwidth]{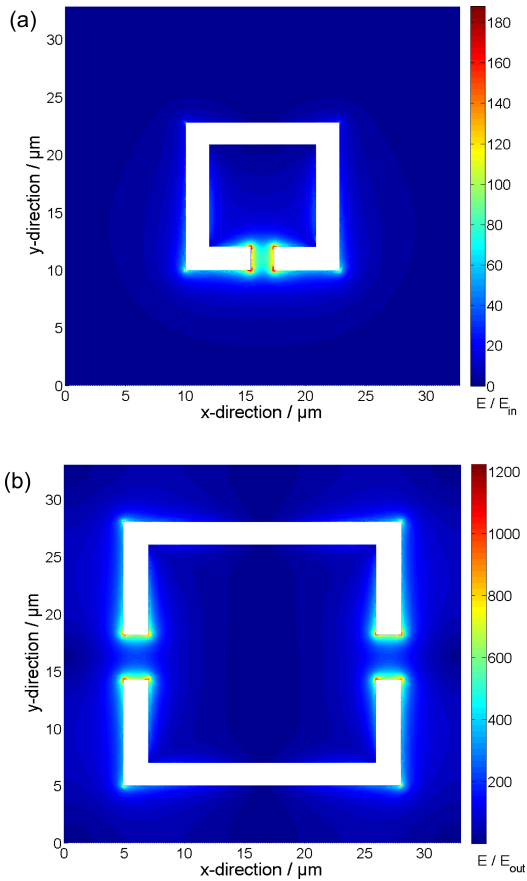}
	\caption{(a) Electric field in the graphene plane for the SRR normalized to the incident THz field. (b) Electric field in the graphene plane for the a-SRR normalized to the radiated field. In both cases the metamaterial/graphene composite consists of  a silver structure and a lossy dielectric with a permittivity of $\epsilon=2$.}
	\label{fig:Efield}
\end{figure}

As can be seen in Fig.\,\ref{fig:Efield}(a) the fields are extraordinarily high in a relatively small area close to the metallic structure. In this respect the estimate that we obtained for the maximally allowed intensity of the incident THz wave describes the worst case since the assumed field enhancement of $g=188$ is only observed in a very small fractional area of the metamaterial unit cell. For example, when we calculate the average field enhancement in the gap we obtain a much more moderate enhancement factor of $g=92$ which yields a maximally allowed intensity of the incident THz wave of $I=900~\mathrm{n W/mm^2}$ for the constant gain regime. Probably the rapid graphene dynamics even allows a much higher THz field intensity for constant gain because local inversion depletion can be readily balanced by high-mobility carriers in the graphene layer such that the average electric field in a unit cell rather than the maximum field or the average field in the split becomes the relevant quantity to describe the inversion dynamics. Taking the average electric field in the unit cell as a basis we obtain an enhancement factor of $g=4.87$ which results in a maximally allowed THz intensity of $300~\mathrm{\mu W/mm^2}$. At this point it should be noted that typical THz field intensities lie in the range of a few microwatts per mm$^2$.

In the case of an a-SRR/graphene composite metamaterial the field enhancement is by far higher than in the case of an SRR/graphene composite, so that the calculated results obtained in the previous chapters are only valid for much lower THz field intensities. Also the goal of using a-SRR/graphene metamaterials is to implement a THz radiation source and the question arises which intensity of emitted THz radiation we can expect. Even though a correct answer requires a more detailed analysis involving a 3-level rate equation model, we can use the numerical results of our 3-D simulations to determine the minimal intensity of radiated THz waves in the constant gain limit $\tilde{n}_{THz}\leq 0.5 \tilde{n}^{abs}_{pump}$. As can be seen in Fig.\,\ref{fig:Efield}(b) the local electric fields are strongly enhanced in the two gaps of the a-SRRs. We determined a maximal local field enhancement of $g=1223$ which results in a THz field intensity of $I_\mathrm{out}\approx 5~\mathrm{nW/mm^2}$. Neglecting the extremely high but locally confined fields by using the average field enhancement in the gap of $g=354$ we obtain a THz field output intensity of $I=60~\mathrm{n W /mm^2}$. Assuming ultrafast carrier dynamics in graphene, as explained before, a much wider intensity range of constant gain can be assumed and we can use the average electric field in the unit cell as responsible quantity for gain. That way we obtain an intensity of $I_\mathrm{out}\approx 1.5~\mathrm{\mu W/mm^2}$ for the radiated THz field when considering a local field enhancement factor of $g=69$. As discussed above, the calculated intensity value are minimal values that we can estimate for operation in the constant gain regime.

We should emphasize that the calculations above serve for the determination of the validity range for constant gain in the metamaterial/graphene composite. This means that we can obtain even higher amplification than in the numerical calculations presented in this paper when we exceed the maximally allowed intensity of the incident THz radiation. However, our numerical calculations lose validity in such a case since the gain in graphene is consequently saturated.
\section{Conclusion}
	We numerically studied amplification of terahertz (THz) radiation in optically pumped split ring resonator (SRR)/graphene and asymmetric split ring resonator (a-SRR)/graphene hybrid systems in the constant gain limit at a temperature of T=77 K and an optical pump wavelength of $1.5~\mathrm{\mu m}$. Employing an analytic model for the complex conductivity of graphene we successively studied the impact of ohmic, dielectric and a combination of ohmic and dielectric loss on the amplification in such metamaterial/graphene composites. Presuming a lossy substrate with a permittivity of $\epsilon=2$ and a loss tangent of $\tan{(\delta)}=0.01$ with graphene and a silver SRR metamaterial on top we obtained an amplified reflectivity of -0.1 dB at a resonance frequency of $2.83~\mathrm{THz}$ and pump intensity of $300~\mathrm{mW/mm^2} $ in comparison with a reflectivity of -3.3 dB in the passive case. This corresponds to a gain of 3.1 dB.
For a corresponding silver a-SRR/graphene composite on the same substrate we obtained an amplified transmission of $32~\mathrm{dB}$ at a resonance frequency of $2.81~\mathrm{THz}$ and a pump intensity of  $300~\mathrm{mW/mm^2}$ in comparison with a transmission of the passive material of $-5~\mathrm{dB}$ dB. This corresponds to a gain of $37~\mathrm{dB}$. For the a-SRR/graphene compositie we can expect the resonators to oscillate in phase. For this reason, a-SRR/graphene composites can act as new sources of coherent THz radiation.
Furthermore, we discussed the limitations of the presumed constant gain restriction and discussed the applicability of our model. It is important to note that our calculations provide a good estimate for the minimum expected amplification in such metmaterial/graphene hybrids and that even higher gain can be implemented when gain saturation is taken into account. 

\section*{Acknowledgment}
P.W. acknowledges financial support by the OPTIMAS Carl-Zeiss-Ph.D. program. J.L.G.-P acknowledges funding by the projects TEC2012-37958-C02-01 and TEC2012-37958-C02-02 and JAE-Doc program, partially supported by ESF.

%
\end{document}